\documentclass[preprint,12pt]{elsarticle}
\usepackage{amssymb}
\journal{Physica E}
\begin{document}

\begin{frontmatter}

\title{Universal Hall conductivity in graphene Maxwell fish-eye quantum dot}

\author{ Zhyrair Gevorkian$^{1,2,*}$ }

\address{$^{1}$ Yerevan Physics Institute,Alikhanian Brothers St. 2,0036 Yerevan, Armenia.\\
$^2$Institute of Radiophysics and Electronics,Ashtarak-2,0203,Armenia}



\begin{abstract}

Graphene quantum dot with Maxwell fish eye potential energy profile is studied. A quasiclassical approximation is used given that potential energy is a slowly varying function of coordinates. Near the zero energy the spectrum of electron  macroscopically degenerates. All the electron trajectories are closed circles that are classified by angular momentum and an additional integral of motion. Provided the complete filling of the lower Dirac zone, a universal value for Hall conductivity  is found.
\end{abstract}
\begin{keyword}
Quantum dot,Maxwell fish-eye,Hall conductivity
\end{keyword}
\end{frontmatter}

\section{Introduction}
Graphene quantum dots (GQD) attract significant interest due to numerous applications \cite{appl}.
One of the most widely-used properties of GQDs is photoluminescence, which is the emission of light by graphene when after absorbtion of the incident photons, excited electrons within the GQD relax back to the lower energy levels.
The Pl properties depend on the energy spectrum of electrons in the GQD. The energy spectrum  itself is determined by the potential energy profile.
Different potential energy profiles for graphene quantum dots were considered  including Coulomb \cite{coulomb}  and other profiles \cite{other}. Main focus of these studies was  the energy spectrum. In the present paper we investigate  the Maxwell fish eye profile which is widely used in optics \cite{born,fish}. It was shown \cite{GDN20} that in this profile in the geometrical approximation limit the closed circular trajectories of the photon can be classified by the integrals of motion. In this paper, we apply these results to the electrons in graphene that are described by the massless Dirac equation.

Different types of Hall currents were considered in graphene.The conventional type of current is associated to the drift of cyclotron orbits of electron in perpendicular magnetic field \cite{graphrev}.

Topological Hall current due to Berry phase and curvature of graphene bands takes place without external magnetic field \cite{gorb2014}. In contrast to conventional case this current is neutral because the contributions of $K^{\prime}$ and $K$ valleys have opposite signs.

In the present paper, we consider another type of Hall current that does not require a magnetic field. It appears because of the drift of circular orbits due to the Maxwell fish-eye potential energy profile. This current originates due to the interaction of the pseudospin of graphene electron and the inhomogeneity.
\section{Model Description}
We start from the Dirac equation for the electrons in graphene \cite{graphrev}
\begin{equation}
 [-iv_F{\bf \sigma}\cdot{\bf \nabla}+(V(r)-E)\hat{I}]\hat{\psi}=0
\label{dirac}
\end{equation}
where ${\bf\sigma}\equiv \sigma_x,\sigma_y$ are two dimensional Pauli matrices, $\hat{I}$ is a unity matrix, $E$ is the energy and $V(r)$ is the external potential that is assumed spherical symmetric. Assume that the potential energy is slow varying function of coordinates. Consequently one can use  quasiclassical approximation and therefore make the substitution $\hat{\psi}=\hat{\psi_p}e^{i{\bf pr}}$
\begin{equation}
[v_F{\bf \sigma}{\bf p}+(V(r)-E)\hat{I}]\hat{\psi_p}=0
\label{quasi}
\end{equation}
Dispersion equation follows from Eq.(\ref{quasi})
\begin{equation}
det[v_F{\bf \sigma}{\bf p}+(V(r)-E)\hat{I}]=0
\label{disp}
\end{equation}
This equation has two solutions describing two branches of the Dirac cone \cite{graphrev}
\begin{eqnarray}
E=-v_Fp+V(r)\\ \nonumber
E=v_Fp+V(r)
\label{twobran}
\end{eqnarray}
If the transitions between two zones are neglected  one can introduce a ``weak zero"\cite{weak} Hamiltonian to describe the processes in each zone separately
\begin{eqnarray}
H_I=-v_Fp+V(r)-E\approx 0\\ \nonumber
\label{twoham}
\end{eqnarray}
For certainty let us consider the lower branch with the first Hamiltonian. We assume that the potential energy profile in quantum dot has the form of Maxwell fish eye within the accuracy of a constant number \cite{born}
\begin{equation}
V(r)-E=\frac{2V_0}{1+\alpha r^2}
\label{eye}
\end{equation}
With this substitution the problem becomes completely analogous to the problem of a photon in an inhomogeneous medium with Maxwell fish eye refraction index profile \cite{born, fish}. We will use all the results already obtained for the photon.It is well-known that \cite{GDN20} in this case along with the angular momentum an additional integral of motion exists
 \begin{eqnarray}
 {\bf L}={\bf r\times p},\quad {\bf T}=(1-\alpha r^2){\bf p}+2\alpha({\bf pr}){\bf r}
 \nonumber \\
 \{{\bf L},H\}=0,\quad \{{\bf T},H\}=0
 \label{intmotion}
 \end{eqnarray}
 where Poisson brackets are determined as
 \begin{equation}
 \{f,g\}=\frac{\partial f}{\partial r_i}\frac{\partial g}{\partial p_i}-\frac{\partial f}{\partial p_i}\frac{\partial g}{\partial r_i}
 \label{Poisson}
 \end{equation}
 As a result of high symmetry, one can be convinced that
\begin{equation}
{\bf T}^2+4\alpha{\bf L}^2=\frac{4V_0^2}{v_F^2}
\label{cas}
\end{equation}
Extended symmetry allows to obtain the ray trajectories without
solving the equations of motion Eq.(\ref{intmotion}). Namely, the vector product of \textbf{$T$} and \textbf{$r$} immediately yields the expression of trajectories,
\begin{equation}
{\bf L}=\frac{{\bf T\times r}}{1-\alpha r^2},\quad
|{\bf r}-{\bf a}|^2=R^2,\quad {\bf a}=\frac{{\bf T\times L}}{2\alpha L^2}, \quad R=\frac{V_0}{\alpha v_F L}
\label{traj}
\end{equation}

As it follows from Eq.(\ref{traj}) the electron trajectories are circles in the $xy$ plane with center  ${\bf a}$ and radius $R$, angular momentum is directed along $z$, ${\bf T}$ lies in the plane \textbf{xy} and ${\bf T}\perp {\bf L}$. It follows from Eq.(\ref{cas}) that there is a trajectory with maximal angular momentum $L_{max}=V_0/\sqrt{\alpha}v_F$. This basic trajectory has the following parameters: ${\bf a}=0$, $r=\alpha^{-1/2}\equiv r_0$, \quad ${\bf pr}=0$, ${\bf T}=0$, \quad $ p_0=L_{max}/r_0=\pm V_0/v_f$. Both signs lead to the same energy $E=-v_F|p_0|+V(r_0)=0$. One can prove that the energy of every trajectory  is also equal to zero. Substituting ${\bf T}$ and ${\bf L}$ into the Casimir condition Eq.(\ref{cas}), one has
\begin{equation}
p=\frac{2p_0}{1+\alpha r^2}
\label{croscas}
\end{equation}
The correspondingly energy on the trajectory is $E=-v_Fp+V(r)=0$. Therefore the state $E=0$ is multiply degenarated (see below). This is analogous to the famous \cite{coudeg} extra $n^2$ degeneracy in Coulomb problem at $E\to 0,\quad n\to\infty$,(see below). To estimate the degeneracy rate note that maximal angular momentum is $p_0r_0$ and the quantization step for angular momentum is $\hbar$.
\begin{equation}
N_d=\frac{2L_{max}}{\hbar}=\frac{2p_0r_0}{\hbar}
\label{deg}
\end{equation}
The coefficient $2$ in Eq.(\ref{deg}) stands for two signs of momentum. Note that in a finite system $V(r)>0$ and therefore the electron energy in the higher graphene zone $E=v_Fp+V(r)$ is always positive and therefore there is a gap between lower and higher zones.
\section{Quantum mechanical consideration}
The Dirac Hamiltonian Eq.(\ref{dirac}) in polar coordinates has the form
\begin{equation}
\hat{H}=\left(
\begin{array}{cc}
V(r),& v_Fe^{-i\varphi}(-i\frac{\partial}{\partial r}-\frac{1}{r}\frac{\partial}{\partial \varphi})\\
v_Fe^{i\varphi}(-i\frac{\partial}{\partial r}+\frac{1}{r}\frac{\partial}{\partial \varphi}),& V(r)\end{array}\right)
\label{polham}
\end{equation}
The wave function equation $\hat{H}\hat{\Psi}=E\hat{\Psi}$ is separable in polar coordinates \cite{sep,per07}. In this case the integral of motion is the total angular momentum $J_z=L_z+\sigma_z/2$ that commutes with the Hamiltonian Eq.(\ref{polham}). The corresponding eigenfunctions are found by the anzatz \cite{sep}
\begin{eqnarray}
\psi_j({\bf r})=\frac{1}{\sqrt{r}}\left(\begin{array}{cc}
         e^{i(j-1/2)\varphi}\Phi^A_j(r)\\
         ie^{i(j+1/2)\varphi}\Phi^B_j(r)\end{array}\right)
 \label{polar}
 \end{eqnarray}
where $j=\pm1/2,\pm3/2...$ and the radial equation reads
\begin{equation}
M_i\Phi_j(r)=0
\label{rad}
\end{equation}
where
\begin{equation}
M_j=\left(\begin{array}{cc}
\varepsilon-V(r)/v_F,&-(\partial_r+j/r)\\
\partial_r-j/r,&\varepsilon-V(r)/v_F\end{array}\right), \quad
\Phi_j\equiv\left(\begin{array}{c}
  \Phi_j^A(r)\\
  \Phi_j^B(r)\end{array}\right)
  \label{radeq}
 \end{equation}
and $\varepsilon\equiv E/v_F$. Separate differential equations of second order  for $\Phi^A_j(\Phi^B_j)$ can be found expressing one function by another from Eq.(\ref{rad}). For simplicity we will assume that $V(r)$ is slowly varying function and the term consisting of its derivative is correspondingly neglected
\begin{eqnarray}
\left[\partial_r^2+\varepsilon^2-2\varepsilon v(r)+v^2-\frac{j(j-1)}{r^2}\right]\Phi_j^A(r)=0 \nonumber\\
\left[\partial_r^2+\varepsilon^2-2\varepsilon v(r)+v^2-\frac{j(j+1)}{r^2}\right]\Phi_j^B(r)=0
\label{abeq}
\end{eqnarray}
where $v(r)\equiv V(r)/v_F$. It follows from Eq.(\ref{abeq}) that the potential energy violates the symmetry between holes and electrons \cite{per07} because the potential energy  term $\varepsilon v(r)$ depends on the sign of $\varepsilon$ and changes the character (attractive or repulsive) for holes and electrons. Another property is that $\Phi^A_j$ can be obtained from $\Phi^B_j$ by substitution $j\to j-1 $. At $\varepsilon=0$ the equation is identical to 3d radial Scr$\ddot{o}$dinger equation for a particle in  attractive potential $-v^2(r)$ \cite{demkov71}. The only difference is that in contrary to the angular momentum  quantum number $l$, the integer values $j$ acquire half-integer values because of the electron pseudospin in graphene. For $\varepsilon=0$ equations Eq.(\ref{abeq}) are exactly solved through the Gegenbauer polynomials \cite{demkov71}. The solution having the correct asymptotics at large distances has the form
\begin{eqnarray}
\Phi_j^{0A}(r)=A2^{-1/2+j}\left(1+r^2/r_0^2\right)^{1/2-j}\left(\frac{r}{r_0}\right)^jC_n^j\left(\frac{r_0^2-r^2}{r_0^2+r^2}\right), \quad j>0\nonumber\\
\Phi_j^{0A}(r)=A_12^{1/2-j}\left(1+r^2/r_0^2\right)^{j-1/2}\left(\frac{r}{r_0}\right)^{1-j}C_n^{-j+1}\left(\frac{r_0^2-r^2}{r_0^2+r^2}\right), \quad j<0
\label{gegen}
\end{eqnarray}
where $n=v_0r_0/2\gg |j|$, $C_n^j$ are the Gegenbauer polynomials \cite{Abstig} , $A,A_1$ are arbitrary constants. The functions $\Phi_j^{0B}(r)$ can be found in analogous manner. Energies close to $0$ can be found from Eq.(\ref{abeq}) substituting wave functions by their values at $\varepsilon=0$, $\Phi_j^{0A,B}$. One can see that Eq.(\ref{abeq}) has two types of solutions: $\varepsilon=0$ and

\begin{equation}
\varepsilon_j=\frac{2\int_0^\infty dr v(r)[\Phi_j^{0A}(r)]^2}{\int_0^\infty dr[\Phi_j^{0A}(r)]^2}
\label{energy}
\end{equation}

By substituting Eq.(\ref{gegen}) into Eq.(\ref{energy}), changing integral variables and using the relation between Gegenbauer and Jakobi polynomials \cite{Abstig} one obtains
\begin{equation}
\varepsilon_j=\frac{2v_0I_{1j}}{I_{2j}}
\label{app}
\end{equation}

where
\begin{eqnarray}
I_{1j}=\int_{-1}^1dy(1-y)^{j-3/2}(1+y)^{j-1/2}[P_n^{(j-1/2,j-1/2)}(y)]^2\nonumber\\
I_{2j}=\int_{-1}^1dy(1-y)^{j-5/2}(1+y)^{j-1/2}[P_n^{(j-1/2,j-1/2)}(y)]^2
\label{app2}
\end{eqnarray}
Here, we assumed $j>0$, however calculations for $j<0$ can be carried out in an analogous manner. At $\varepsilon=1/2,3/2$  integrals diverge, however the divergence of the second integral is stronger therefore $\varepsilon_{(1/2,3/2)}=0$. For the other values of $j$ we evaluate integrals numerically. The results are shown in Fig.1
 \begin{figure}
\begin{center}
 \vspace{1cm}
\includegraphics[width=8.0cm]{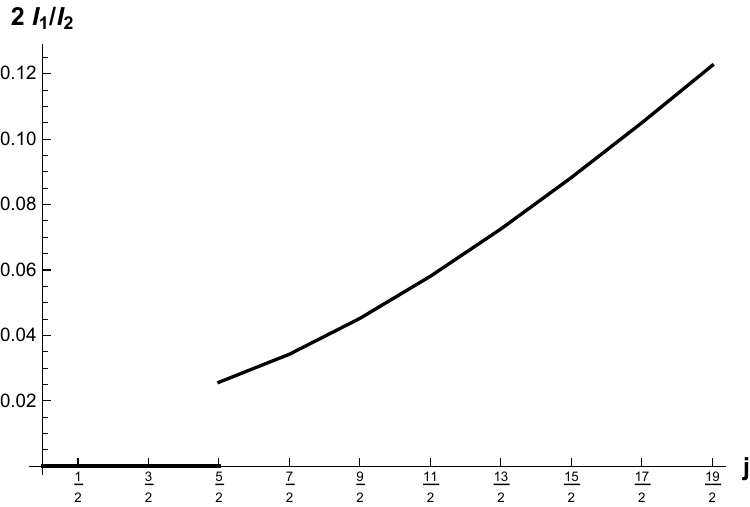}
\caption{Energy dependence on orbital quantum number $j>0$. For $j<0$, one has a symmetrical picture.}
\label{fig.1}
\end{center}
\end{figure}

As it is seen from Fig.1 there is a gap between macroscopically degenerated $\varepsilon=0$ state and non-degenerate positive energy states.In the next section we will consider the kinetic properties of the system provided that Fermi energy lies in this gap.

\section{Kinetics}
It is obvious from the previous section that quantum consideration leads to the same result\textbf{,} namely the macroscopical degeneracy of the state $E=0$ . Quantum Hamiltonian will originate other deeper $E<0$ states that our quasiclassical approach does not capture. However, they will not contribute to the linear response of the graphene electron gas to the static electric field provided that Fermi energy is at $E_F=0$.

The current density is determined as
\begin{equation}
{\bf j}(t)=-\frac{4e}{S}\sum_{{\bf p}}\int\frac{ d{\bf r}}{4\pi r_0^2}{\bf \dot{r}}f({\bf r},{\bf p},t)
\label{equation}
\end{equation}
Here the coefficient $4$ stands for spin and valley degeneracy, $f$ is the distribution function that should be found from the kinetic equation.
The equilibrium distribution function at $t=-\infty$ has the form
\begin{equation}
f({\bf r,p},t=-\infty)=\Theta(E_f+v_Fp-V(r))
\label{equil}
\end{equation}

Here we assume that the Fermi energy lies in the lower branch and thus do not consider the transitions between the Dirac cones.
Afterwards, according to the linear response theory, one adiabatically switches on the electrical field that reaches its constant value at $t$ to induce the current.
The equations of motion have the form
\begin{equation}
{\bf \dot{r}}=-v_F\frac{{\bf p}}{p},\quad {\bf \dot{p}}=-\frac{\partial V}{\partial {\bf r}}+e{\bf E}(t)
\label{eqmot}
\end{equation}

To construct the distribution function evolving in time we should note that it explicitly depends on ${\bf r}$, $t$ as variables and it also depends on ${\bf p=p(t=-\infty)}$ as a parameter. ${\bf p}(t)=\int_{-\infty}^td\tau[-\frac{\partial V}{\partial{\bf r}(\tau)}+e{\bf E}(\tau)]+{\bf p}$ following from the equation of motion Eq.(\ref{eqmot}) contributes to the $t$ dependence of $f$.

The distribution function obeys the kinetic equation that in the collisionless case has the form
\begin{equation}
\frac{\partial f}{\partial t}+\frac{\partial f}{\partial {\bf r}}{\bf \dot{r}}=0
\label{kin}
\end{equation}
 The solution of  Eq.(\ref{kin}) obeying the boundary conditions Eq.(\ref{equil}), ${\bf r}(t)\equiv {\bf r}$, ${\bf p}(t=-\infty)\equiv {\bf p}$ has the form
\begin{equation}
f({\bf r}, {\bf p}(t),t)=\Theta\left(v_F|{\bf p}(t)|-V(r)+e{\bf E(t) r}-e\int_{-\infty}^t d\tau {\bf r}(\tau){\bf\dot{E}(\tau)}\right)
\label{sol}
\end{equation}
where Fermi energy is assumed to be equal $E_F=0$.
To find conductivity one should expand the distribution function and keep linear terms of ${\bf E}$
\begin{equation}
{\bf j}(t)=-\frac{4e^2}{S}\sum_{\bf p}\int\frac{d{\bf r}}{4\pi r_0^2}{\bf \dot{r}}\delta\left[v_Fp(t)-V(r)\right]\left[{\bf E(t) r}-\int_{-\infty}^t d\tau {\bf r}(\tau){\bf\dot{E}(\tau)}\right]
\label{lincur}
\end{equation}
Assuming that external electric field is directed along $y$, integrating by parts and using the equations of motion Eq.(\ref{eqmot}), one has
\begin{equation}
{\bf j}(t)=\frac{4e^2v_F}{S}\sum_{\bf p}\int\frac{d{\bf r}}{4\pi r_0^2}\frac{{\bf p(t)}}{p(t)}\delta\left[v_Fp(t)-V(r)\right]\int_{-\infty}^tE(\tau)dy(\tau)
\label{part}
\end{equation}
In ${\bf p}(t)$ in Eq.(\ref{part}), electric field in linear approximation can be put $E(t)=0$ . Therefore the unperturbed circles Eq.(\ref{traj}) that ensure the zeros of the argument of delta function can give  contribution to the integral on ${\bf r}$. Each circle has its counterpart that differs by the sign of $y$ and these opposite circles cancel each others contribution to ${\bf j}$, see Fig.1.  Therefore the only contribution to the integral comes from the basic trajectory (see above) with the center at ${\bf a}=0$ and radius $r\equiv r_0$ that has not a counterpart, see Fig.1.

 \begin{figure}
\begin{center}
 \vspace{1cm}
\includegraphics[width=8.0cm]{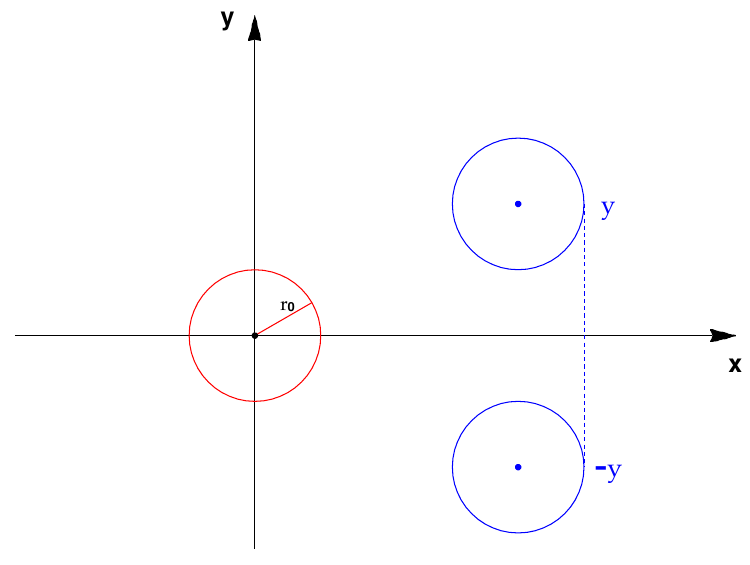}
\caption{The only contribution to Hall conductivity gives basic trajectory(red), contributions from paired(blue) trajectories cancel to each other.}
\label{fig.2}
\end{center}
\end{figure}

Hence taking into account the property of delta function with complex function in argument, the current density can be rewritten
\begin{equation}
{\bf j}(t)=\frac{4e^2v_F}{S}\sum_{\bf p}\int\frac{d{\bf r}}{4\pi r_0^2}\frac{{\bf p_0(t)}}{p_0(t)}\frac{\delta(r-r_0)}{|\frac{\partial V}{\partial r}|}\int_{-\infty}^tE(\tau)dy_0(\tau)
\label{rewrit}
\end{equation}
 Subscript zero means the coordinates and momentums on the basic trajectory. The current density Eq.(\ref{rewrit}) is the instantaneous  value at a given time $t$. However, the physical meaning has the averaged over a period of motion in a basic circle. Before that by taking into account that $E(\tau)$ is slowly varying compared the period function, one can substitute $\int_{-\infty}^tE(\tau)dy_0(\tau)=Ey_0(t)$, where $E\equiv E(t)=const$ and we assume that $y_0(-\infty)=0$. The average current density over the period takes the form
\begin{equation}
<{\bf j}(t)>=\frac{4e^2v_F E}{S}\sum_{\bf p}\int\frac{d{\bf r}}{4\pi r_0^2}<\frac{{\bf p_0(t)}y_0(t)}{p_0(t)}>\frac{\delta(r-r_0)}{|\frac{\partial V}{\partial r}|}
\label{average}
\end{equation}

In order to carry out the averaging over a period, remind the time dependencies of coordinates and momenta when the electrons are moving over a circle of radius $r_0$
\begin{eqnarray}
x_0=r_0\sin\omega t,\quad y_0=r_0\cos\omega t,\quad \omega=\frac{v_F}{r_0}\nonumber\\
p_x=p_0\cos\omega t,\quad p_y=-p_0\sin\omega t,\quad p_0=V_0/v_F
\label{tdep}
\end{eqnarray}
It follows from Eq.(\ref{tdep}) that only the cross term average gives a nonzero contribution
\begin{equation}
<p_x(t)y_0(t)>=p_0r_0<\cos^2\omega t>=\frac{p_0r_0}{2}
\label{avtime}
\end{equation}
Substituting Eq.(\ref{avtime}) into Eq.(\ref{average}) and taking the integral, for the Hall conductivity one has
\begin{equation}
\sigma_{xy}=\frac{e^2v_F r_0}{V_0}\frac{1}{S}\sum_{\bf p}
\label{Hall}
\end{equation}
The quantity $N=\sum_{\bf p}$ consists of two parts: the number of all states at zero energy i.e. the degeneracy rate Eq.(\ref{deg}) and the number of ${\bf p}$ within a closed trajectory with given angular momentum $N=N_d N_{\bf p}$. This number we calculate for the basic trajectory with maximal angular momentum $p_0r_0$
\begin{equation}
N_{\bf p}=S\int \frac{d{\bf p}}{(2\pi)^2}\delta(p^2r_0^2-p_0^2r_0^2)=\frac{S}{4\pi r_0^2}
\label{enpi}
\end{equation}
Substituting Eq.(\ref{deg}) and Eq.(\ref{enpi}) into Eq.(\ref{Hall}), we finally obtain
\begin{equation}
\sigma_{xy}=\frac{e^2}{h}
\label{final}
\end{equation}
In the same conditions of graphene electron gas one, can be convinced that diagonal terms of conductivity tensor are equal to zero: $\sigma_{xx}=\sigma_{yy}=0$. The appearance of nonzero Hall conductivity in the absence of external magnetic field is a result of interaction of graphene electron pseudospin and inhomogeneity.  The latter is analogous to spin Hall effects in optics \cite{spinhallop} and condensed matter \cite{spinhallcond}.
\section{Summary}
We have considered the spectrum and kinetic properties of a graphene quantum dot with Maxwell fish-eye potential energy profile. The quasiclassical approximation is used which is correct provided the electron de Broglie wavelength is much smaller than the characteristic radius of the potential energy profile $\hbar v_F/V_0\ll r_0$. The problem is reduced to the motion of a massless particle in a Maxwell fish-eye potential energy profile. Because of the existence of an additional integral of motion, the circular trajectories of the particle  are found without solving the equations of motion. It is shown that the $E=0$ state is macroscopically degenerated. Hall conductivity acquires a universal value $\sigma_H=e^2/h$ provided that Fermi energy  is zero.


\subsection{Acknowledgements}

This work was supported by the Armenian Science Committee
projects 20RF-023 and 21AG-1C062.

\end{document}